\documentclass[reprint,aps,pra]{revtex4-1}

\usepackage{graphicx}
\usepackage{dcolumn}
\usepackage{bm}
\usepackage{amsmath}
\usepackage{amsthm}
\usepackage{amssymb}
\usepackage{float}
\usepackage{hyperref}
\usepackage{color}
\usepackage{epstopdf}
\usepackage{cleveref}
\usepackage[svgnames]{xcolor}
\hypersetup{hidelinks,colorlinks=true,allcolors=DarkBlue}

\newcommand{\ket}[1]{|#1\rangle}

\newcommand{\Rx}[2]{{R}_x^{(#1)}\left(#2\right)}
\newcommand{\Ry}[2]{{R}_y^{(#1)}\left(#2\right)}

\begin{document}

\preprint{APS/123-QED}

\title{Multilevel superconducting circuits as two-qubit systems: \\ Operations, state preparation, and entropic inequalities}
\author{E.\,O.\,Kiktenko$^{1,2}$}
\author{A.\,K.\,Fedorov$^{3,1,*}$}
\author{O.\,V.\,Man'ko$^{4}$}
\author{V.\,I.\,Man'ko$^{4}$}
\affiliation
{
\mbox{$^{1}$Bauman Moscow State Technical University, Moscow 105005, Russia}
\mbox{$^{2}$Geoelectromagnetic Research Center of Schmidt Institute of Physics of the Earth,}
\mbox{Russian Academy of Sciences, Troitsk, Moscow Region 142190, Russia}
\mbox{$^{3}$Russian Quantum Center, Skolkovo, Moscow 143025, Russia}
\mbox{$^{4}$P.\,N. Lebedev Physical Institute, Russian Academy of Sciences, Moscow 119991, Russia}
}

\date{\today}

\begin{abstract}
We theoretically study operations with a four-level superconducting circuit as a two-qubit system. 
Using a mapping on a two-qubit system, we show how to implement iSWAP gates and Hadamard gates through pulses on transitions between particular pairs of energy levels. 
Our approach allows one to prepare pure two-qubit entangled states with desired form of reduced density matrices of the same purity and, in particular, arbitrary identical reduced states of qubits. 
We propose using schemes for the Hadamard gate and two-qubit entangled states with identical reduced density matrices in order to verify $\log{N}$ inequalities for Shannon and R\'enyi entropies for the considered noncomposite quantum system.

\begin{description}
\item[PACS numbers]
03.65.Wj, 03.65.-w
\end{description}
\end{abstract}
                              
\maketitle

\section{Introduction}

During last decade, quantum correlations and the entanglement phenomenon in composite quantum systems, {\it i.e.}, systems containing subsystems, has been viewed as a potential resource for technologies  
such as high-efficiency information processing \cite{Ladd}, 
long-distance secure communications \cite{Gisin}, 
ultra-sensitive metrology \cite{Ye}, 
simulation of complex systems \cite{Cirac}, 
and many others.
Inspiring progress on control for quantum systems on the level of individual quantum objects, {\it e.g.}, 
in experiments with 
photons, 
electrons, 
nitrogen-vacancy centers, 
nuclear spins, 
quantum dots, 
optomechanical systems,
superconducting circuits, 
ultracold trapped atoms, ions and polar molecules, has been achieved. 
However, the development of a universal toolbox for efficient control for large quantum systems scalable with respect to number of subsystems
is still a crucial challenge of quantum science and technologies \cite{Lukin}. 

Fundamental results on generalization of the Shannon classical information theory in the quantum domain have been obtained \cite{Shannon,Holevo,Nielsen}. 
Quantum correlations for states of composite systems are successfully described in terms of various information and entropic characteristics, 
including the von Neumann entropy and quantum mutual information \cite{Neumann}, 
discord related measures \cite{discord,discord2,discord3}, 
informational asymmetry \cite{causaltiy},
contextuality \cite{Manko2},  
entropic inequalities \cite{inequalities,inequalities2,inequalities3},
subadditivity and strong subadditivity conditions \cite{inequalities3,Petz,Lieb,MAManko1}. 
Entropic characteristics of quantum states have been widely studied \cite{Petz,MAManko1,MAManko2} in the framework of $q$-deformed entropic functions, 
{\it e.g.}, R\'enyi \cite{Renyi} and Tsallis entropies \cite{Tsallis}, 
depending on single extra parameter, as well as a larger number of parameters \cite{Hu}. 

Recently, possibilities of using noncomposite ({\it i.e.}, indivisible) quantum systems as a potential platform for test of underlying principles of quantum physics \cite{Zeilinger,MAManko3,MAManko4,MAManko5}
and realizations of quantum technologies have been discussed \cite{MAManko3,MAManko4,MAManko5,Kessel,Kessel2,Paraoanu}.
This paradigm dates back to the foundation of the Kochen--Specker theorem \cite{KochenSpecker}, which provides certain type of constraints on hidden variable theories, that could be used to explain probability distributions of quantum measurement outcomes.
In particular, a photonic qutrit ($d=3$) has been used for experimental demonstration of fundamentally non-classical properties of noncomposite quantum systems \cite{Zeilinger}.

\begin{figure}[t] 
\begin{centering}
\includegraphics[width=0.3\textwidth]{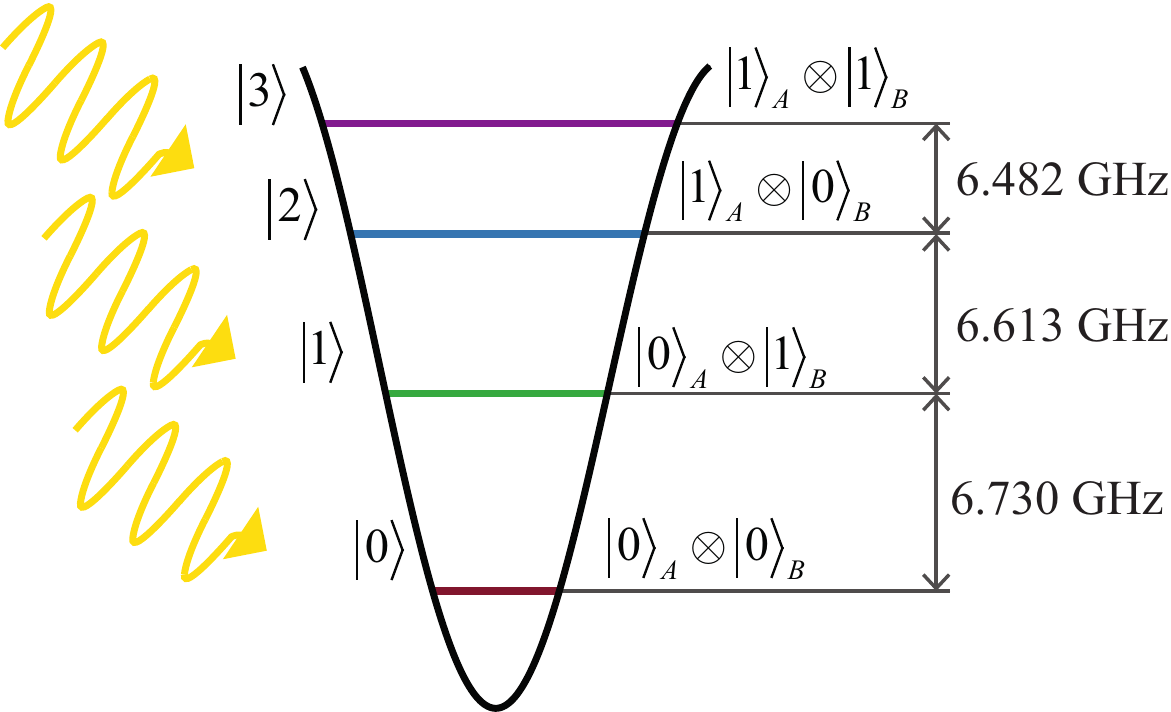}
\end{centering}
\vskip -4mm
\caption
{
Isomorphic correspondence between a qudit state with $d{=}4$ and a two-qubit system. 
Transition frequencies between levels correspond to the multilevel superconducting circuit investigated in Ref. \cite{Katz3}.
}
\label{fig:main}
\end{figure}

For a $j=3/2$--spin system, information and entropic characteristics have been analyzed in detail \cite{MAManko3,Kessel,Kessel2,Markovich}. 
The information properties of quantum states and their characteristics could be associated indeed with both composite and noncomposite systems.
Then all the above mentioned information and entropic measures for composite quantum systems can be mapped in a very simple way on the case of noncomposite quantum systems 
\cite{MAManko3,MAManko4,MAManko5,Kessel,Paraoanu}.
This framework opens up new perspectives for the implementation of quantum technologies, {\it e.g.}, 
many-level quantum simulation \cite{Paraoanu} and quantum information processing \cite{Kessel,Kessel2,Kiktenko}, with noncomposite quantum systems.

In the present work, we report about operation with a four-level superconducting circuit as a two-qubit system.
We consider the implementation of various quantum gates and preparation of a specific class of two-qubit states.
In particular, we show how to realize iSWAP gate and Hadamard gate through applying pulses on the transition between particular pairs of energy levels in the system.
We introduce a scheme for preparation of two-qubit pure entangled states with desired reduced density matrices of the same purity, as well as states with identical matrices. 
Combining these approaches, we demonstrate how these states can be used for verification of entropic inequalities, related with the quantity $\log{N}$ with $N$ being the dimensionality of the Hilbert space \cite{logn}, for noncomposite systems. 
In short, we denote this class as $\log{N}$ entropic inequalities.
This type of inequalities is a particular case of the Massen--Uffink entropic uncertainty relation \cite{logn2}, which are in their turn a fundamental aspect of quantum physics.
Another important feature of these inequalities is a link to the quantum Fourier transform \cite{MAManko2}, which plays a crucial role for several quantum algorithms, {\it e.g.}, the Shor's algorithm \cite{Shor}.

In the present proposal, 
the noncomposite system is implemented as a single multilevel quantum system realized via anharmonic superconducting circuit (see Fig.~\ref{fig:main}).
Superconducting circuits with Josephson junctions have been considered in early works \cite{DodonovManko1,Manko4,OVManko4,Dodonov1,Dodonov2,DodonovLozovik}
as well as in recent theoretical \cite{Zeilinger2,Zeilinger3,OVManko2,Wallraff,You} and experimental studies 
\cite{Wallraff,You,Delsing,Delsing2,Abdumalikov,Wallraff2,Hakonen,Hakonen2,Delsing3,Hakonen3,Katz,Katz2,Katz3,Gustavsson,Ustinov,DiCarlo,Martinis,Wallraff3,Obada,Wallraff4}.
Noncomposite quantum systems can be realized through a variety of physical platforms, which include, but are not limited to photons \cite{Padua} and NMR systems \cite{Gedik}.
Nevertheless, progress in experiments with multilevel systems based on superconducting circuits forces us to focus our study on this environment 
\cite{Delsing,Delsing2,Wallraff2,Hakonen,Hakonen2,Abdumalikov,Delsing3,Hakonen3,Katz,Katz2,Gustavsson,Katz3,Ustinov}. 

Superconducting circuits indeed can be considered as highly tunable artificial atoms. 
Being a very useful environment for demonstration of analogs of phenomena related to interactions between atoms and electromagnetic radiation such as 
dynamical Casimir effect \cite{Delsing,Delsing2}, 
Autler-Townes splitting \cite{Wallraff2,Hakonen,Hakonen2}, 
electromagnetically induced transparency \cite{Abdumalikov}, and dynamical Lamb effect \cite{Lozovik}, 
these systems are promising candidates for quantum computing \cite{DiCarlo,Martinis,Wallraff3,Obada} and simulation \cite{Wallraff4}.
Performance of two-qubit quantum algorithms \cite{DiCarlo,Martinis} as well as implementations of a three-qubit operation have been shown \cite{Wallraff3}.
Schemes for realizations of the quantum Fourier transform have been considered \cite{Obada}.

In the recent experimental study \cite{Katz3}, isomorphic correspondence between a four-level qudit and a two-qubit quantum system has been used to explore ``hidden'' two-qubit dynamics of a four-level superconducting circuit with the Josephson junction. 
In fact, this straightforwardly demonstrates a potential resource of noncomposite systems realized as multilevel superconducting circuits for quantum technologies. 

Quantum correlations and quantum discord phenomena for coupled $LC$-nanoelectric circuits \cite{Kiktenko2} and superconducting circuits with the Josephson junction \cite{Yu} have been discussed.
However, the details of entropic inequalities for the Shannon entropy and the R\'enyi entropy and a possible way for their verification can be clarified. 

Our paper is organized as follows.
In Section \ref{system}, 
we describe an isomorphic correspondence between the qudit with $d=4$ and two-qubit quantum system as well as demonstrate possible schemes for a set of quantum gates for our noncomposite system: iSWAP and Hadamard gates.
We briefly discuss extension of the system towards realization of the universal set of two-qubit gates.
In Section \ref{preparation}, we introduce a scheme for the preparation of two-qubit entangled states with equal levels of purity and, in particular, states with identical reduced density matrices. 
In this framework, we discuss the application of such schemes aimed on verification of particular cases of entropic inequalities in noncomposite quantum systems based on anharmonic superconducting circuits.
Finally, in Section \ref{conclusions} we summarize the results of our work.

\section{Multilevel quantum system: \\ Quantum gates}\label{system}

Our approach can be applied to any many-level quantum system.
Here, we focus on the realization with superconducting anharmonic multi-level circuits due to significant experimental progress in this scope \cite{Katz,Katz2,Gustavsson,Katz3,Ustinov}.

\subsection{Mapping on a bipartite quantum state}

In this work, we are particularly interested in the realization of a four-level quantum system. 
One can introduce the following mapping of an ``original'' four-level system on an ``artificial'' bipartite (two-qubit) system \cite{MAManko3,MAManko4,MAManko5,Kessel} as the following isomorphic correspondence between the stationary energy states and two-qubit logic basis:
\begin{equation}\label{eq:map}
	|0\rangle\leftrightarrow|00\rangle, \quad |1\rangle\leftrightarrow|01\rangle, \quad |2\rangle\leftrightarrow|10\rangle, \quad |3\rangle\leftrightarrow|11\rangle,
\end{equation}
where we use the following notation (see Fig. \ref{fig:main}):
$$
	\ket{ij}\equiv\ket{i}_A\otimes\ket{j}_B.
$$ 
Here, $A$ and $B$ stand for qubits of an artificial two-qubit quantum system.

\begin{figure*}[t]
\begin{centering}
\includegraphics[width=0.975\textwidth]{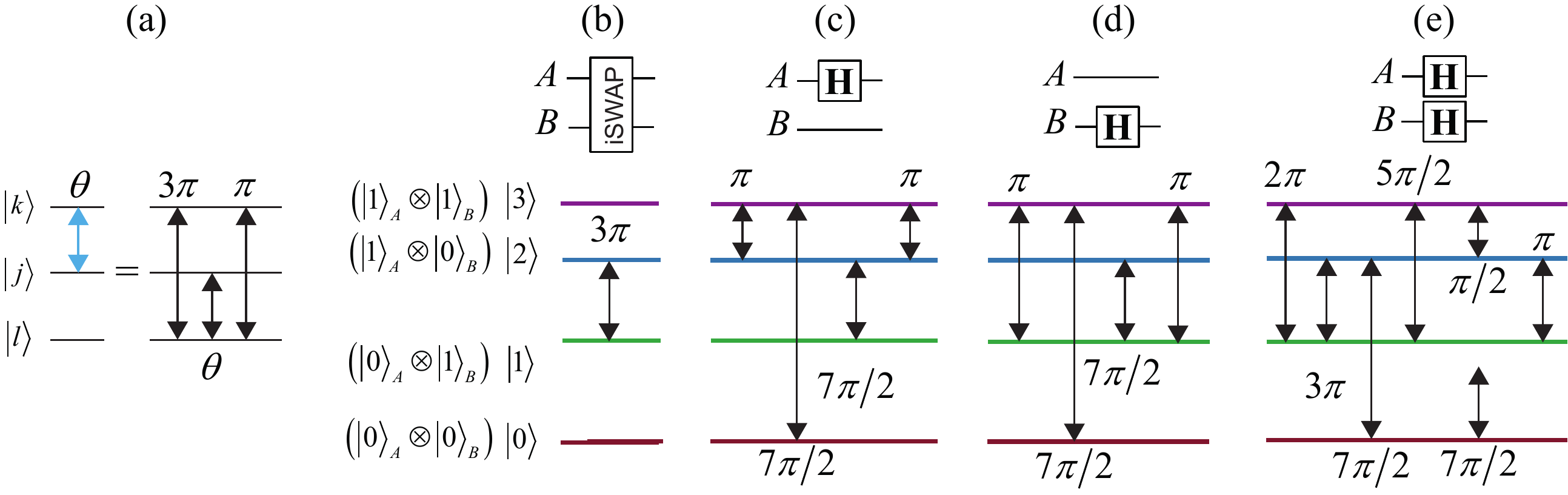}
\end{centering}
\vskip -2mm
\caption
{
In (a) realization of rotation around the $y$ axis in the Bloch sphere of  the corresponding two-dimensional Hilbert subspace spanned by vectors $\ket{j}$ and $\ket{k}$ (blue arrow) 
using only operators of rotations around the $x$ axis (black arrows) applied to different couples of three-level set $\{\ket{j}, \ket{k}, \ket{l}\}$ [see~Eq. (\ref{eq:Ry})].
In (b)--(e) we show the implementation of various gates via sequences of $\theta$ pulses (\ref{eq:thetapulse}) applied to different two-level transitions. 
The duration of each pulse is denoted by the value near the corresponding arrow.
We note that in the figure, time is directed from left to right, {\it i.e.}, the order of pulses is opposite to that in Eqs.~(\ref{eq:Ry}), (\ref{eq:iSWAP}), (\ref{eq:HadA})--(\ref{eq:HadAB}).
}
\label{fig:2}
\end{figure*}

For the original quantum system with the density matrix written in the computational basis
\begin{equation} \label{eq:rho0}
	\rho=\rho_{AB}=
	\begin{bmatrix}
		\rho_{00} & \rho_{01} & \rho_{02} & \rho_{03} \\
		\rho_{01}^{*} & \rho_{11} & \rho_{12} & \rho_{13} \\
		\rho_{02}^{*} & \rho_{12}^{*} & \rho_{22} & \rho_{23} \\
		\rho_{03}^{*} & \rho_{13}^{*} & \rho_{23}^{*} & \rho_{33} \\
	\end{bmatrix}
\end{equation}
we obtain the following density matrices of qubits in their computational basis:
\begin{equation}\label{eq:rhoA}
\begin{aligned}
	\rho_A&=\begin{bmatrix}
		\rho_{00}+\rho_{11} & \rho_{02}+\rho_{13} \\
		\rho_{02}^{*}+\rho_{13}^{*} & \rho_{22}+\rho_{33}
	\end{bmatrix}, \\
	\rho_B&=\begin{bmatrix}
		\rho_{00}+\rho_{22} & \rho_{01}+\rho_{23} \\
		\rho_{01}^{*}+\rho_{23}^{*} & \rho_{11}+\rho_{33}
	\end{bmatrix}.	 
\end{aligned}
\end{equation}

We note that there are other suitable ways to map multilevel systems on two-qubit systems such as ``Bell-state'' encoding \cite{Katz3}.
However, it is crucial that the diagonal elements of qubit matrices (\ref{eq:rhoA}) are composed of the diagonal elements of density matrix (\ref{eq:rho0}).
This feature is the result of using mapping (\ref{eq:map}).

In this case, we find that the measurement of the original system state in the quanta number basis is equivalent to the simultaneous measurements of both artificial subsystems in their computational basis.
Realization of unitary rotation operators previous to such measurement opens a way to the tomographic characterization of qubit states.

\subsection{Quantum logic gates}

We consider a possible scheme for a set of quantum logic gates for our noncomposite system.
We assume that the only operation we can perform is applying a $\theta$ pulse on the transition between any pair of four energy levels in the system.
It is realizable by coupling of a superconducting circuit to an external resonant field \cite{You}.
Non-linearity of the potential, which comes from the Josephson junction, makes it possible to address a particular transition in the system at least on a theoretical level of consideration.

The corresponding operator of $\theta$ pulse, which is a rotation operator around $x$ axis in the Bloch sphere of the corresponding two-dimensional Hilbert subspace, reads
\begin{equation}\label{eq:thetapulse}
	R_x^{(jk)}(\theta)= \begin{bmatrix}
		\cos({\theta}/{2}) & -i\sin({\theta}/{2}) \\
		-i\sin({\theta}/{2}) & \cos(\theta/{2}) 
	\end{bmatrix}^{(jk)}
	\oplus
	{\mathbb I}^{(\overline{jk})},
\end{equation}
where the superscript $j,k\in\{0,1,2,3\}$ of the matrix indicates that it is written in the basis $\{\ket{j},\ket{k}\}$, 
$\oplus$ denotes the direct sum, 
and ${\mathbb I}^{\overline{(jk)}}$ stands for the identity operator acting in the orthogonal complement $\left(\mathrm{Span}\{\ket{j},\ket{k}\}\right)^\bot$, 
such that the operator ${R}_x^{(jk)}(\theta)$ acts in the whole Hilbert space of the four-level system.

The useful feature of the considered system is that the proper sequence of pulses, 
originally corresponded to rotation around $x$ axis, 
allows one to implement the effective rotation around the $y$ axis of the particular Bloch sphere [see Fig. \ref{fig:2}(a)]:
\begin{equation}
\begin{split}\label{eq:Ry}
	R_y^{(jk)}(\theta)&=\begin{bmatrix}
		\cos({\theta}/{2}) & -\sin({\theta}/{2}) \\
		\sin({\theta}/{2}) & \cos(\theta/{2}) 
	\end{bmatrix}^{(jk)}
	\oplus
	{\mathbb I}^{(\overline{jk})}=\\
	&=R_x^{(jl)}(\pi)R_x^{(kl)}(\theta)R_x^{(jl)}(3\pi).
\end{split}
\end{equation}

We note that in~(\ref{eq:Ry}) the ancillary energy level $l$ from $\overline{jk}$ is used. 
Its occupation and phase in the result of this sequence of pulses remains intact.

Using (\ref{eq:thetapulse}) and (\ref{eq:Ry}) allows us to construct important logical gates for our noncomposite system.
As an illustrative example, we consider the two-qubit iSWAP gate, which can be realized as follows [see also Fig. \ref{fig:2}(b)]:
\begin{equation} \label{eq:iSWAP}
\begin{split}
U_\mathrm{iSWAP}\equiv
	\begin{bmatrix}
		1 & 0 & 0 & 0 \\
		0 & 0 & i & 0 \\
		0 & i & 0 & 0 \\
		0 & 0 & 0 & 1
	\end{bmatrix}
	&={{R}_x^{(12)}(3\pi)}= \\ 
	&=\begin{bmatrix}
		0 & i \\
		i & 0 
	\end{bmatrix}^{(12)}
	\oplus
	{\mathbb I}^{(03)},
\end{split}
\end{equation}
where the the full matrix is written in the computational basis.
One can see that its realization is rather straightforward, because only a single appropriate rotation operation is needed. 

One-qubit operations, in contrast, appear to be more intricate for implementation.
Here, we focus on the realization of the Hadamard gate
\begin{equation} \label{eq:HadG}
	{\bf H}
	=\frac{1}{\sqrt{2}}
	\begin{bmatrix}
		1 & 1\\
		1 & -1
	\end{bmatrix}.
\end{equation}

The application of the Hadamard gate to both qubits of the Hadamard gate (\ref{eq:HadG}) to particular qubits $A$ and $B$ corresponds to the following sequences of the rotation operators:
\begin{eqnarray}\label{eq:HadA}
\begin{split}
	&{\bf H}^{(A)}\otimes{\mathbb I}^{(B)}=
	\frac{1}{\sqrt{2}}
		\begin{bmatrix}
		1 & 0 & 1 & 0 \\
		0 & 1 & 0 & 1 \\
		1 & 0 & -1 & 0 \\
		0 & 1 & 0 & -1
	\end{bmatrix}=\\
	&={R}_y^{(13)}\left(\frac{\pi}{2}\right){R}_y^{(02)}\left(\frac{\pi}{2}\right){R}^{(23)}_x(2\pi)=\\
	&=\Rx{23}{\pi}\Rx{12}{\frac{7\pi}{2}}\Rx{03}{\frac{7\pi}{2}}\Rx{23}{\pi},
\end{split} \\
\label{eq:HadB}
\begin{split}
	&{\mathbb I}^{(A)}\otimes{\bf H}^{(B)} =
	\frac{1}{\sqrt{2}}
		\begin{bmatrix}
			1 & 1 & 0 & 0 \\
			1 & -1 & 0 & 0 \\
			0 & 0 & 1 & 1 \\
			0 & 0 & 1 & -1
		\end{bmatrix}=\\
		&={R}_y^{(23)}\left(\frac{\pi}{2}\right){R}_y^{(01)}\left(\frac{\pi}{2}\right){R}^{(13)}_x(2\pi)=\\
		&=\Rx{13}{\pi}\Rx{12}{\frac{7\pi}{2}}\Rx{03}{\frac{7\pi}{2}}\Rx{13}{\pi},
\end{split}
\end{eqnarray}
where superscripts $(A)$ and $(B)$ indicate that the particular operator acts in the Hilbert subspace of the corresponding qubit [see Figs. \ref{fig:2}(c)--(d)].

The application of the Hadamard gate (\ref{eq:HadG}) to the both qubits can be realized by the coherent implementation of (\ref{eq:HadA}) and (\ref{eq:HadB}), which needs eight pulses, or via the following optimized sequence:
\begin{equation}\label{eq:HadAB}
\begin{split}
	&{\bf H}^{(A)} \otimes{\bf H}^{(B)} =
		\frac{1}{2}
		\begin{bmatrix}
			1 & 1 & 1 & 1 \\
			1 & -1 & 1 & -1 \\
			1 & 1 & -1 & -1 \\
			1 & -1 & -1 & 1
		\end{bmatrix}=
		\\
	&={R}_y^{(13)}\left(\frac{\pi}{2}\right){R}_y^{(02)}\left(\frac{\pi}{2}\right){R}_y^{(23)}\left(\frac{5\pi}{2}\right)\times
	\\&~~~~~~~~~~~~~~~~~~\times{R}_y^{(01)}\left(\frac{\pi}{2}\right){R}^{(13)}_x(2\pi)=\\
	&=
	\Rx{12}{\pi}\Rx{23}{\frac{\pi}{2}}
	\Rx{01}{\frac{7\pi}{2}}\Rx{13}{\frac{5\pi}{2}}\times\\
	&~~~~~~~~~~~~~~~~~~\times\Rx{02}{\frac{7\pi}{2}}\Rx{12}{3\pi}\Rx{13}{2\pi},
\end{split}
\end{equation}
which uses seven pulses [see Fig.\ref{fig:2}(e)]. 

It is important to note that using such methods does not allow to realize the universal set of two-qubit quantum gates \cite{Nielsen}: one-qubit Hadamard gate, $\pi/8$-gates, and controlled NOT gate.
Indeed, our primary rotation operators (\ref{eq:thetapulse}) and (\ref{eq:Ry}) together with implemented logic gates given by Eqs. (\ref{eq:iSWAP}) and (\ref{eq:HadA})--(\ref{eq:HadAB}) belong to the SU$(4)$ group.
Apparently, one can not realize any operator beyond the SU$(4)$ group by a sequence of rotation operator (\ref{eq:thetapulse}).

A particular example of such operator is the controlled-NOT gate,
\begin{equation}
	U_\mathrm{CNOT}=\begin{bmatrix}
		1 & 0 & 0 & 0 \\
		0 & 1 & 0 & 0 \\
		0 & 0 & 0 & 1 \\
		0 & 0 & 1 & 0
	\end{bmatrix},
\end{equation}
which has the determinant being equal to $-1$.

Thus, its implementation requires another approach for implementation and realization of noncomposite systems.
A possible way towards overcome this challenge is using ancillary level is for realization of operators from U$(4)$ group through operators from SU$(5)$ group.
This strategy has been recently studied in the context of the Deutsch algorithm realization using five-level anharmonic superconducting multilevel quantum circuits \cite{Kiktenko}. 

\section{Operating with two-qubit states}\label{preparation}

Here, we consider an example operation with four-level superconducting circuit as two-qubit system aimed on preparation pure two qubit entangled state with desired reduced density matrices of the same purity.
As a particular case, we demonstrate how to prepare the state with identical reduced matrices.
Such class of two-qubit states can be used for verification of $\log{N}$ entropic inequalities.

\subsection{Preparation of states with equal purity}

We assume that our four-level circuit is initially in the ground state $\ket{0}$.
Our aim is to prepare a two-qubit state $\ket{\psi}$ with desired reduced density matrices $\rho_A$ and $\rho_B$ having the equal modules of corresponding Bloch vectors.
This equality is a consequence of the purity of the joint state, which implies the equal purity of marginals:
\begin{equation}
	\mathrm{Tr}\rho_A^2=\mathrm{Tr}\rho_B^2.
\end{equation}

\begin{figure}[t]
\begin{centering}
\includegraphics[width=1\linewidth]{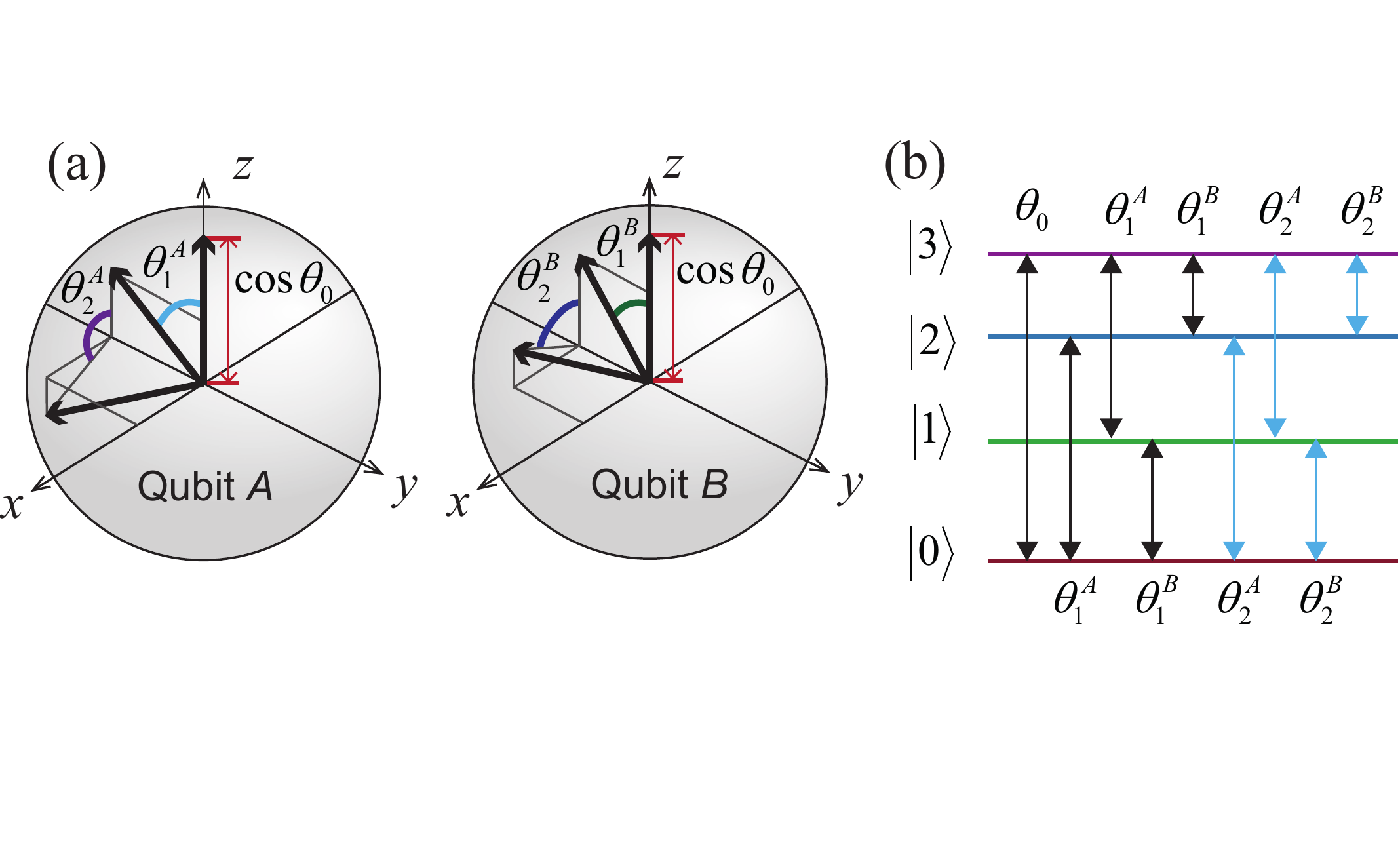}
\end{centering}
\vskip -15mm
\caption
{
The suggested scheme of preparation a pure two-qubit state $\ket{\psi}$ (\ref{eq:psi}) with reduced states having the same purity (length of Bloch radius vector): 
In (a) reduced states of qubits in Bloch sphere representation. 
In (b) sequence of pulses applied to four-level superconducting circuit that prepares $\ket{\psi}$ from the ground state $\ket{0}$, where the black arrows stands for $x$ rotations~(\ref{eq:thetapulse}) and blue for $y$ rotations~(\ref{eq:Ry}).
} 
\label{fig:3}
\end{figure}

One can cope with preparation of the state $\ket{\psi}$ in three steps.
The first step is to apply entangling pulse $\Rx{03}{\theta_0}$ which sets radii of the Bloch vectors of $\rho_A$ and $\rho_B$ to $\cos\theta_0$ (see Fig.~\ref{fig:3}).

The second is to apply rotations around $x$ axis on angles $\theta_1^A$ and $\theta_2^B$ in the Bloch spheres of the both qubits Hilbert spaces:
\begin{equation}
	\begin{aligned}
		R_x^{(A)}(\theta_2^A)=\Rx{13}{\theta_2^A}\Rx{02}{\theta_2^A}, \\
		R_x^{(B)}(\theta_2^B)=\Rx{01}{\theta_2^B}\Rx{23}{\theta_2^B}. \\
	\end{aligned}
\end{equation}

Finally, the third step is to apply rotations around the $y$ axis for the both qubits as follows:
\begin{equation}
	\begin{aligned}
		R_y^{(A)}(\theta_3)=\Ry{13}{\theta_2^A}\Ry{02}{\theta_2^A}, \\
		R_y^{(B)}(\theta_3)=\Ry{01}{\theta_2^B}\Ry{23}{\theta_2^B}. \\
	\end{aligned}
\end{equation}

As a result, one can obtain the following state:
\begin{multline}
\label{eq:psi}
	\ket{\psi}=R_y^{(B)}(\theta_2^B)R_y^{(B)}(\theta_2^A)R_x^{(B)}(\theta_1^B)R_x^{(A)}(\theta_1^A)\times\\
	\times\Rx{03}{\theta_0}\ket{0},
\end{multline}
with reduced states
\begin{equation} \label{eq:rho}
	\rho_j=\begin{bmatrix}
	1+z_j & x_j+iy_j \\
	x_j-iy_j & 1-z_j
\end{bmatrix},
\end{equation}
where the following notations are used:
\begin{equation}
	\begin{aligned}
		&x_j=\cos{\theta_0}\cos{\theta_1^j}\sin{\theta_2^j},\\
		&y_j=-\cos{\theta_0}\sin{\theta_1^j}, \\
		&z_j=\cos{\theta_0}\cos{\theta_1^j}\cos{\theta_2^j}, \quad j=A,B.
	\end{aligned}
\end{equation}

Thus, using of proper sequence of $\theta$ pulses allows to prepare arbitrary reduced qubit states of the same purity.

\subsection{States with identical reduced density matrices}

There is an important particular case of the discussed above scheme, which allows to prepare two-qubit states  with identical reduced matrices:
\begin{equation}
	\rho_A=\rho_B=\rho_0.
\end{equation} 

Towards this end, we consider a simple modification of this scheme, with setting in Eq. (\ref{eq:psi}):
\begin{equation}
	\theta_1^A=\theta_1^B, 
	\quad 
	\theta_2^A=\theta_2^B.
\end{equation}

In reality, the identity between resulting states of qubits depends strongly on the accuracy of applied pulses.

\subsection{Application: verification of entropic inequalities}
We note that the class of two-qubit states with identical reduced density matrices can be used for verification of particular types of entropic inequalities, being manifestation of uncertainty relations written in terms of information theory.
The general idea is to construct a two two-qubit state, with reduced density matrix of one qubit being Hadamard transform of the other one, and then study their properties.

For instance, one can prepare an entangled state with identical reduced matrices $\rho_0$, apply Hadamard gate on the second qubit $B$, perform a measurement in computational basis of the whole system getting the diagonal elements of the each qubit, 
and verify the following $\log 2$ inequality~\cite{logn}, being a particular case of Maassen-Uffink uncertainty relation~\cite{logn2}:
\begin{equation}\label{eq:ineq_1}
	H(\rho_0)+H({\bf H}\rho_0{\bf H})\geq\log{2},
\end{equation}
where
\begin{equation} \label{eq:SE}
	H(\sigma)=-\sum_{m}{\sigma_{mm}\log\sigma_{mm}}
\end{equation}
is the Shannon entropy calculated from measurement statistic in computational basis.

The similar inequality
\begin{equation}\label{eq:ineq_2}
	\mathcal{R}_{\alpha}(\rho)+\mathcal{R}_{\beta}({\bf H}\rho{\bf H})\geq\log{N}, \qquad \alpha^{-1}+\beta^{-1}=2
\end{equation}
can be considered for R\'enyi entropy
\begin{equation} \label{eq:RE}
	\mathcal{R}_{\alpha}(\rho)=\frac{1}{1-\alpha}\log\sum_{m}({\rho_{mm})^\alpha}, \quad \alpha\ge0,
\end{equation}
which reduces to Shannon entropy at $\alpha=1$.

We should note two points. 
First, the verification of inequalities (\ref{eq:ineq_1}) and (\ref{eq:ineq_2}) requires that the reduced operators were identical to a very high degree of accuracy.
Second, the considered setup of four-level circuit also allows to check $\log 4$ inequalities by preparing some pure two-qubit state $\ket{\psi}$, measuring it in computational basis, then preparing the state again, applying Hadamard gates to the both qubits obtaining vector
\begin{equation} 
	{\bf H}^{(A)}\otimes{\bf H}^{(B)}\ket{\psi},
\end{equation}
and also measuring it in the computational basis.

The limitation of the last scheme is that we are restricted to the space of pure two-qubit spaces with assuming, that our circuit is initially in the ground state $\ket{0}$ and we can neglect decoherence processes during operation with the system.  

\section{Conclusions and outlook}\label{conclusions}

We summarize the main results of the present paper. 
Using the isomorphic correspondence between qudit with $d=4$ and ``two-qubit'' quantum system, we considered possible schemes for a set of quantum logic gates for our noncomposite system: iSWAP and Hadamard gates.
We discussed possible methods for producing classes of two-qubit states with the same purity and the same reduced density matrices.
We suggested to use these schemes for verification of $\log{N}$ inequalities for Shannon and R\'enyi entropies, which are one of manifestation of uncertainty relations for conjugate variables. 

A link between our consideration and the framework of highly controllable and easily implementable in experiment platform results in opportunities of the verification of $\log{N}$ entropic inequalities via existing experimental tools \cite{Katz,Katz2,Gustavsson,Katz3,Ustinov}, 
and, furthermore, consider possibilities to investigate computational speed-up in oracle-based quantum algorithms with the use of multilevel artificial atoms.

\section*{Acknowledgments}
We thank S. Filippov, K. Shulga, and E. Glushkov for fruitful discussions and useful comments.   
This work is supported by the RFBR, the Dynasty Foundation, and the Council for Grants of the President of the Russian Federation (Grant No. SP-961.2013.5).

\vspace{\baselineskip}

*Corresponding author: {\href{mailto:akf@rqc.ru}{akf@rqc.ru}}

\end{document}